\documentclass[seceq]{ptptex}

\usepackage{graphicx}
\usepackage{epsf}
\usepackage{wrapft}

\def\beq{\begin{equation}}
\def\eeq{\end{equation}}

\notypesetlogo                       

\title{Structure of Metastable States in Phase Transitions with High-Spin Low-Spin Degree of Freedom}

\author{Seiji {\sc MIYASHITA}$^{1)}$\footnote{E-mail address: miyashita@phys.s.u-tokyo.ac.jp},
Yusuke {\sc KONISHI}$^{1)}$\footnote{E-mail address: konishi@spin.phys.s.u-tokyo.ac.jp},
Hiroko {\sc TOKORO}$^{1)}$\footnote{E-mail address: tokoro@spin.phys.s.u-tokyo.ac.jp},
Masamichi {\sc NISHINO}$^{2)}$\footnote{E-mail address: NISHINO.Masamichi@nims.go.jp},
Kamel {\sc BOUKHEDDADEN}$^{3)}$\footnote{E-mail address: kbo@physique.uvsq.fr}
 and 
Francios {\sc VARRET}$^{3)}$\footnote{E-mail address: varret@physique.uvsq.fr}
}

\inst{$^1$ Department of Physics, School of Science, The University of Tokyo, Hongo, Tokyo 113-8656, Tokyo 113-8656\\
$^2$Computational Materials Science center, National Institute for Materials Science, Tsukuba 305-0047, Japan\\
$^3$Laboratoire de Magnetisme et d~AfOptique, Universite de Versailles Saint-Quentin-en-Yvelines, 
45 Avenue des Etats Unis 78035 Versailles.
}

\abst{
Difference of degeneracy of the low-spin (LS) and high-spin (HS) states causes 
interesting entropy effects on spin-crossover phase transitions 
and charge transfer phase transitions in materials composed of the spin-crossover atoms.
Mechanisms of the spin-crossover (SC) phase transitions have been studied by 
using Wajnflasz model, where the degeneracy of the spin states (HS or LS) is 
taken into account and cooperative natures of
the spin-crossover phase transitions have been well described.
Recently, a charge transfer (CT) phase transition due to electron hopping between LS and HS sites
has been studied by using a generalized Wajnflasz model.
In the both systems of SC and CT, 
the systems have a high temperature structure (HT) and a low temperature
structure (LT), and the  change between them can be a smooth crossover or a discontinuous 
first order phase transition depending on the parameters of the systems.
Although apparently the standard SC system and the CT system are very different, 
it is shown that both models are equivalent under a certain transformation of variables.
In both systems, the structure of metastable state at low temperatures is a matter of interest.
We study temperature dependence of fraction of HT systematically
in a unified model, and
find several structures of equilibrium and metastable states of the model 
as functions of system parameters. 
In particular, we find a reentrant type metastable branch of HT in a low temperature region,
which would play an important role to study the photo-irradiated processes of 
related materials.
} 
  
\begin{document}

\maketitle

\section{Introduction}

In the so-called spin-crossover atoms, e.g., Fe and Co, spin states of
the atoms can be changed between
the high-spin (HS) state and the low-spin (LS) state by small perturbations, such as
changes of temperature and pressure and also photo-irradiation, etc.
This change of spin state is called
spin-crossover (SC) transition.\cite{sce1,sce2,sce3,sce4,sce5,sce6,sce7,sce8,sce9,sce10,sce11,sce12,sce13,sce14,sce14.1,sce14.2,sce15,sce16,sce17,sce18,sct1,sct2,sct3,sct3.1,sct4,sct5,sct6,sct7,sct8,sct9,sct10,sct11,sct12,sct13,sct16,sct17,sct18,sct19,sct20,wm1,wm2} 
The HS state has a large number of degeneracy and it is favorable at high temperatures, while
the LS state has low energy which is favorable at low temperatures.
The competition between the effects of degeneracy (the entropy) and of the energy 
provides interesting changes of structure of material.

From a view point of adiabatic energy levels of local structure of atoms, 
the population dynamics among states (HS, LS, and other excited states) under photo-irradiation 
has been studied by a kind of rate equation, 
which explains mechanism of transition among the LS and HS states.\cite{sce14.1,sce14.2}
It has been also pointed out that cooperative interactions are important for SC transitions.
With an aid of interaction between atoms, this transition can be either smooth crossover or 
discontinuous first order phase transition depending on the system parameters.
\cite{sce1,sce2,sce3,sce4,sct1,sct2,sct3,sct3.1,sct4,sct5,sct6,sct7,sct8,sct9,sct10,sct11,sct12,sct13,wm1,wm2}
This situation is well described by the Wajnflasz model.\cite{wm1,wm2}
Control between the HS and LS states has been realized by photo-irradiation as 
the light-induced excited spin state trapping (LIESST),
\cite{sce13,sce14,sce14.1,sce14.2,sce15,sce16,sce17,sce18,sct16,sct17,sct18,sct19,sct20}
and structure of the metastable ordered state of the systems 
has become an important topic.\cite{sce18,sct21,pba18,pba19,pba20}

Prussian blue analogues (PBAs), classified as molecule-based 
magnets,\cite{pba1,pba2,pba3,pba4,pba5,pba6,pba7,pba8} 
show various photomagnetic 
phenomena.\cite{pba9,pba10,pba11,pba12,pba13,pba14,pba15,pba16,pba17}
Two-way photo-switching between magnetic and nonmagnetic states 
was observed in a Co-Fe Prussian blue analogue,
K$_{0.2}$Co$_{1.4}$Fe(CN)$_6$6.9H$_2$O\cite{pba10,pba11}, which has attract much 
interest. 
This photomagnetism is based on the charge-transfer-induced spin
transition (CTIST) between two phases,\cite{pba13} 
e.g., one is the high temperature (HT) phase consisting mainly of the Co$^{\rm II}$-Fe$^{\rm III}$
where Co$^{\rm II}$ is in the high spin state (HS, $S$ = 3/2) and 
Fe$^{\rm III}$ is in the low spin state (LS, $S$ = 1/2),
and the other is the low temperature (LT) phase consisting mainly of the Co$^{\rm III}$-Fe$^{\rm II}$ where
Co$^{\rm III}$ is in the low spin state (LS, $S$ = 0) and Fe$^{\rm II}$ 
is also in the low spin state (LS, $S$ = 0). 
In this material, Co and Fe ions
are antiferromagnetically coupled. 
Therefore, precisely speaking, it is a ferrimagnetic material. 
However, the spin-crossover transition occurs at the Co site, and 
if we take a unit of the pair atoms (Fe and Co), 
we may regard the transition as a generalized SC transition.

On the other hand, another type of phase transition due to the charge transfer (CT) 
was discovered in a mixed-valence iron complex 
(n-C$_{3}$H$ _{7}$)$_{4}$N[Fe$^{\rm{II}}$Fe$^{\rm{III}}$(dto)$_{3}$]
(dto = C$_{2}$O$_{2} $S$_{2}$) \cite{ct1,ct2,ct3}. 
In this material, the charge transfer causes a
change of degeneracy of atomic states, and a phase transition occurs between the high
temperature (HT) and low temperature (LT) structures. 
In this system, the spin transition on each atom does not occur. 
However, the entropy effect has an important role for
the structure change of electron configuration 
and nature of phase transition was explained by using a generalized Wajnflasz model.\cite{ct3}
There, it was found that a metastable branch of HT exists at all the temperatures 
below the critical temperature.
This behavior is qualitatively different from that in the standard model of SC transition. 
That is, because experimentally the fraction of HS shows a hysteresis loop in SC transitions,
for SC transitions we have adopted a model where 
the metastable branch of HS terminates at a spinodal point, and below which the 
HS is unstable. 
In this model, it is difficult to explain the metastable behavior of
photo-induced HS state at low temperatures. 
In order to overcome this difficulty, a dynamical effect has been introduced by
adopting the so-called Arrhenius dynamics~\cite{sct16}, which has explained successfully
the fact that the HS state remains for a long time at low temperatures although it is
unstable in a mean-field free energy. \cite{sct18,sct19,sct20}
On the other hand, in the model of CT, the HT state is metastable at all the temperatures.
Now, it is an interesting problem to study the relation between the both models. 

In this paper, we will show that the both models are equivalent under a certain
transformation of variable, and the unified model shows several qualitatively different 
structures of equilibrium and metastable states as a function of system parameters.
In particular, a reentrant type metastable branch of HT in a low temperature region
is discovered. The existence of this low temperature metastable structure
would play an important role to study the photo-irradiated processes of
related materials.
Here, we mainly study the transition of spin states, e.g. HS $\Leftrightarrow$ LS. 
However, if we consider magnetic interaction
between the spins, the structure of phase transition is modified. 
The magnetic effect has been studied for Co-Fe PBA.
\cite{sct19,sct20,pba18,pba19}  
The magnetic phase transition in the CT system has been also studied.\cite{ct3,ct4,ct5} 
The existence of the metastable states causes new variety of magnetic phase transition,
which will be briefly discussed.

\section{Unified model}

\subsection{Typical spin-crossover phase transition with a single hysteresis}

In order to describe 
the spin-crossover transition, the so-called Wajnflasz model has been adopted,\cite{wm1,wm2} where
the HS state is represented by $s=1$, and the LS state is represented by $s=-1$.
It should be noted that
the states $s_i=\pm 1$ have different degeneracy. 
Let the number of states of the HS state be $n_{\rm H}$, and that of the LS state $n_{\rm L}$.
Let us review the Wajnflasz model briefly.\cite{sct16}
The interaction among atoms is originated in the elastic interaction in the
atomic structure.\cite{sct4} 
However, here it is simply modeled by a nearest-neighbor interaction between sites,
and the Hamitonian has the following form
\beq
{\cal H}_{\rm W}=-J\sum_{<ij>}s_is_j+D\sum_i s_i,
\label{SCs}
\eeq
where $J$ denotes the interaction between states of neighboring atoms, 
and $D$ denotes the energy difference between HS and LS. 
Here we consider the case where $D>0$, i.e., the LS state is energetically favorable.
The partition function is given by
\beq
Z={\rm Tr}e^{-\beta{\cal H}_{\rm W}}
=\sum_{s_1=\pm 1}\ '\cdots\sum_{s_N=\pm 1}\ '
\exp\left({\beta J\sum_{<ij>}s_is_j-\beta D\sum_i s_i}\right),
\eeq
where $\sum'$ denotes that the  summation is carried out over the degenerate states, i.e., 
summation over $n_{\rm H}$ states of $s_i=1$ and $n_{\rm L}$ states of $s_i=-1$. 
Here, $\beta = 1/k_{B}T$, and $N$ is the number of sites.
This form is expressed by a non-degenerate Ising variable $\sigma_i=\pm 1$ as
\beq
Z
=\sum_{\sigma_1=\pm 1}\cdots\sum_{\sigma_N=\pm 1}\exp\left({\beta J\sum_{<ij>}\sigma_i\sigma_j-\beta D\sum_i \sigma_i}\right)
\prod_i\left(n_{\rm H}\delta_{\sigma,1}+n_{\rm L}\delta_{\sigma,-1}\right).
\eeq
We may rewrite the second factor as
\beq
n_{\rm H}\delta_{\sigma,1}+n_{\rm L}\delta_{\sigma,-1}=\sqrt{n_{\rm H}n_{\rm L}}
e^{{1\over2}\sigma \ln {n_{\rm H}\over n_{\rm L}}}.
\label{degeneracy}
\eeq
Now we have
\beq
Z =(n_{\rm H}n_{\rm L})^{N\over2}\sum_{\sigma_1=\pm 1}\cdots\sum_{\sigma_N=\pm 1}
\exp\left(\beta \left(J\sum_{<ij>}\sigma_i\sigma_j
-\left(D-{k_{\rm B}T\over2}\ln{g}\right)\sum_i\sigma_i\right)\right),
\label{W0-model}
\eeq
and 
\beq
g={n_{\rm H}\over n_{\rm L}}.
\eeq
Therefore the model (\ref{SCs}) can be expressed by an effective Hamiltonian with
a temperature dependent field
\beq
{\cal H}=-J\sum_{<ij>}\sigma_i\sigma_j
+\left(D-{k_{\rm B}T\over2}\ln{g}\right)\sum_i\sigma_i.
\label{W-model}
\eeq
This form of Hamiltonian is called Wajnflasz model.

Here, it should be noted on a characteristic property of this model.
At high temperatures, $k_{\rm B}T > 2D/\ln g$, the term of the effective field 
($H_{\rm eff}=-D+k_{\rm B}T\ln g/2$)
is positive, and thus the spins has a positive expectation value, $\langle \sigma_i\rangle >0$.
On the other hand, at low temperatures,  $k_{\rm B}T < 2D/\ln g$,  $\langle \sigma_i\rangle < 0$.
Let us define the marginal temperature $T_0$ to be
\beq
k_{\rm B}T_0={2D\over \ln g},
\label{T0}
\eeq
where $\langle \sigma_i\rangle = 0$.
If $T_0$ is larger than the critical temperature $T_{\rm IC}$ of the corresponding Ising model
\beq
{\cal H}=-J\sum_{<ij>}\sigma_i\sigma_j,
\label{Ising0}
\eeq
that is,
\beq
T_0> T_{\rm IC},
\label{smooth}
\eeq
then the change of the magnetization is smooth.   
On the other hand, if
\beq
T_0 < T_{\rm IC},
\label{discontinuous}
\eeq
then a discontinuous change occurs, which means the first order phase transition takes place.

\begin{figure}
$$
\begin{array}{c}
\epsfxsize=7.0cm \epsfysize=6.5cm \epsfbox{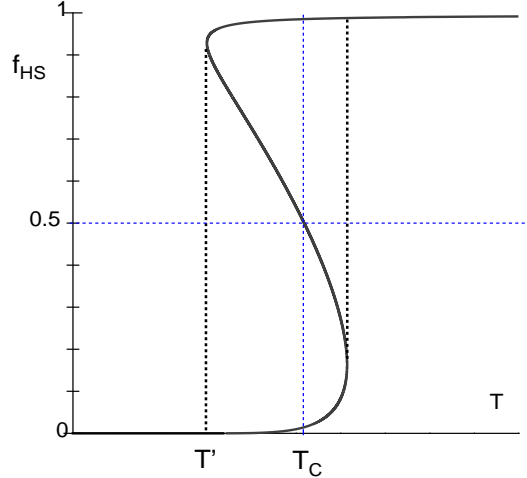}
\end{array}
$$
\caption{Schematic temperature dependence of a solution of the self-consistent equation of a
mean-field theory for the high spin fraction $q$ of 
models of SC transition with a hysteresis loop.  
The bold dotted curves denote the hysteresis region. 
The thin dotted line denotes the first order phase transition temperature where $q=0.5$.} 
\label{fig-hysteresis}
\end{figure}
A typical example of the hysteresis obtained by a kind of mean field analysis 
is schematically depicted in Fig.~\ref{fig-hysteresis}.
So far, the experimental results of the first order phase transition have been
explained by this type of temperature dependence of order parameter $f_{\rm HS}$
$(=(\langle{\sigma _i}\rangle +1)/2)$ which is the
fraction of the atoms in the HS state.
In this figure, the HS state becomes unstable at a temperature $T'$
denoted by a dotted line. Below this temperature, the HS state is unstable.
Thus, it seems difficult to have a long-lived HS state at low temperatures.
In Co-Fe PBA, however, the magnetic state exists for a long time after pumping by photo-irradiation.
In order to explain this long-lived state we have introduced a mechanism of very slow dynamics,
i.e., the so-called Arrhenius dynamics\cite{sct16,sct20}, where we assume an 
Arrhenius type relaxation time: $\tau\propto \exp(-E_0/k_{\rm B}T)$ with an appropriate
value of $E_0$. 

\subsection{Metastable structure of a charge transfer phase transition}

\begin{figure}
$$
\begin{array}{ccc}
\epsfxsize=4.1cm \epsfysize=4.5cm \epsfbox{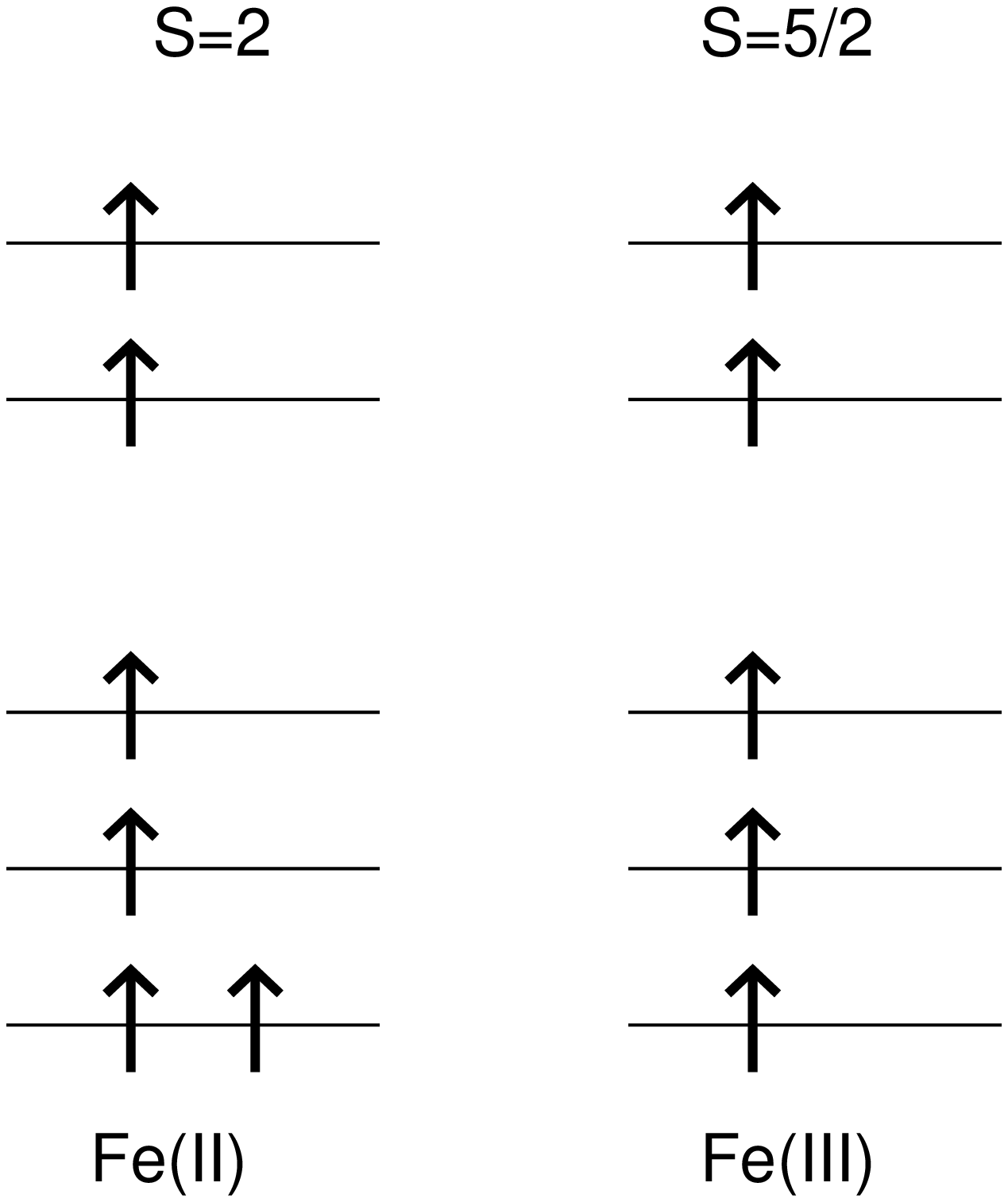}&\hspace{10mm}&
\epsfxsize=4.1cm \epsfysize=4.5cm \epsfbox{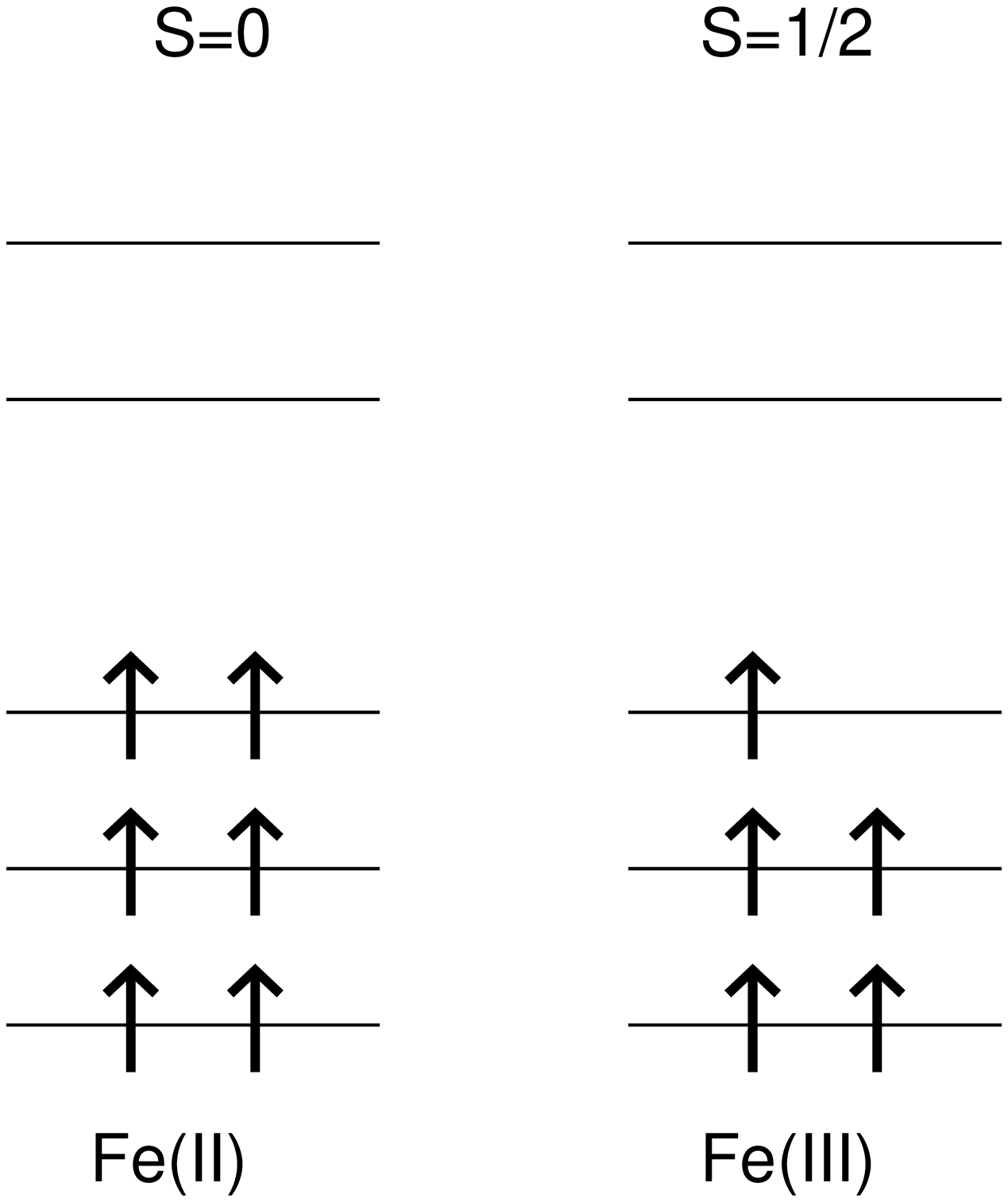}\\
({\rm a}) \quad {\rm High \ spin \ (HS) \ states}  
&& ({\rm b}) \quad {\rm Low \ spin \ (LS)\  states}
\end{array}
$$
\caption{Electron states of Fe$^{\rm{II}}$ and Fe$^{\rm{III}}$ 
(a) in the high spin (HS) state and (b) in the low spin (LS)state} 
\label{fig-Fe}
\end{figure}

Contrary to the above mentioned behavior, 
we have found a case where the static metastability exists 
in the study of the charge transfer transition in 
material  
(nC$_{3}$H$_{7}$)$_{4}$N[Fe$^{\rm{II}}$Fe$^{\rm{III}}$(dto)$_{3}$]
(dto=C$_{2}$O$_{2} $S$_{2}$)\cite{ct1,ct2,ct3}.
Here, let us consider this charge transfer transition. 
This material consists of bipartite lattice. 
At one of the sublattices, Fe is surrounded by sulphur atoms where Fe is always in 
the low spin state (LS), and 
at the other sublattice Fe is surrounded by oxygen atoms where Fe is always in the high spin state (HS).
We call the former 'A-site' and the latter 'B-site'. 
Let the both sublattices have $N$ sites. 
Fe$^{\rm{II}}$ has one more electron than Fe$^{\rm{III}}$, 
and the difference between them can be
expressed by the number of additional electron $n_i$ which is 1 in Fe$^{\rm{II}}$ and 0 in Fe$^{\rm{III}}$.
Here, we consider the degeneracy of spin degree of freedom.\cite{ct3} 
At A-site, Fe$^{\rm{II}}$ is in the low spin state and $S=0$ as depicted in Fig.~\ref{fig-Fe} 
and thus the degeneracy is 1, and 
the spin of Fe$^{\rm{III}}$ is $S=1/2$ and thus its degeneracy is 2.
Similarly at B-site, $S=2$ for Fe$^{\rm{II}}$ and $S=5/2$ for Fe$^{\rm{III}}$.
In terms of $\{n_i\}$, the degeneracy is given as the following: 
\begin{center}
\begin{tabular}{cccc} 
site & $n$ & spin $S$ & degeneracy\\
\hline
A  & $n=1$ & 0   &$n_{\rm A1}$=1 \\  
A  & $n=0$ & 1/2 &$n_{\rm A0}$=2 \\  
B  & $n=1$ & 2   &$n_{\rm B1}$=5 \\  
B  & $n=0$ & 5/2 &$n_{\rm B0}$=6 \\
\hline 
\end{tabular}
\end{center}
\vspace{5mm}

\begin{figure}
$$
\begin{array}{ccc}
\epsfxsize=7.0cm \epsfysize=6.5cm \epsfbox{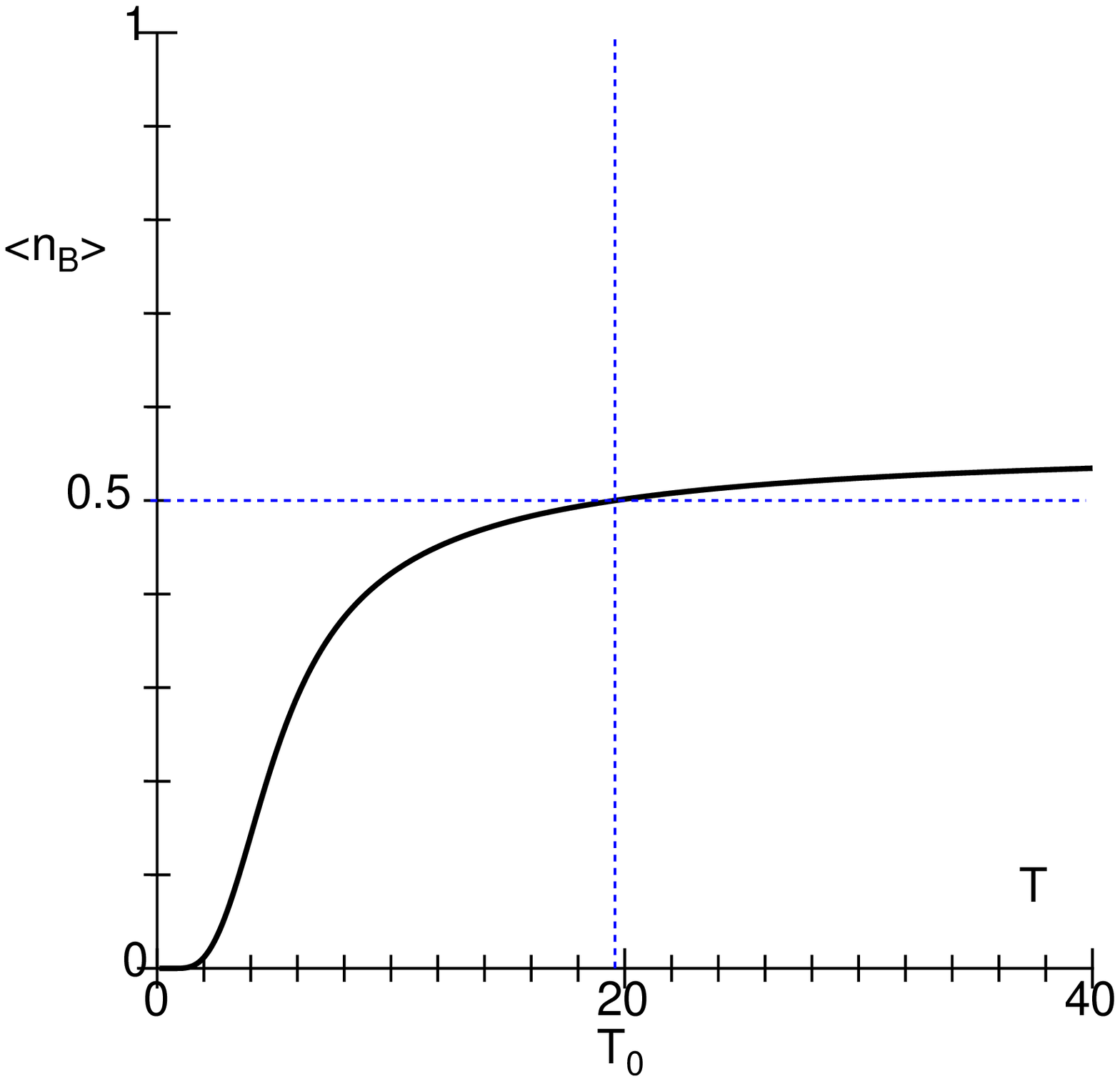}&
&
\epsfxsize=7.0cm \epsfysize=6.5cm \epsfbox{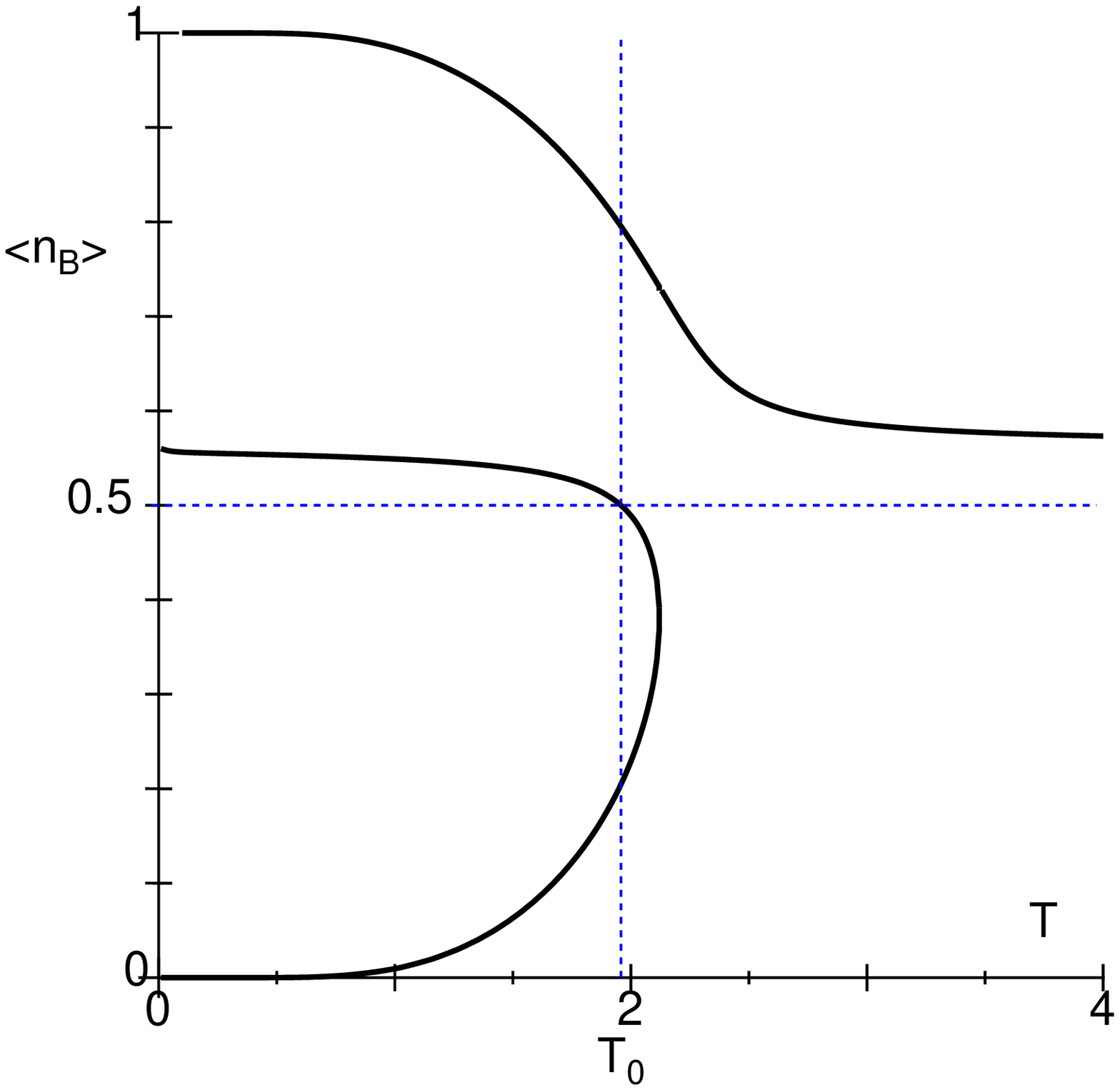}\\
({\rm a}) && (\rm b)
\end{array}
$$
\caption{Temperature dependence of fraction of the HT state 
$\langle n_{\rm B}\rangle$ for systems with $\varepsilon=$ 3. (a) $\Delta=10$ and (b) $\Delta=1$.
$T_0$ is the temperature where $\langle n_{\rm B}\rangle=0.5$ and also it is the critical temperature 
in the case (b).}  
\label{fig-CTnB}
\end{figure}

\noindent
Because the numbers of sites of Fe$^{\rm{II}}$ and Fe$^{\rm{III}}$
are the same, the total number of the additional electrons is $N$
\beq
\sum_in_i=N.
\eeq
In this material, electrons transfer between A and B sites.
We have introduced the following Hamiltonian for this system
\beq
{\cal H}_{\rm CT}=\varepsilon\sum_{<ij>}n_in_j+\Delta\sum_{i\in {\rm B}} n_i,
\label{HamCT}
\eeq
where $\varepsilon$ represents the interaction between the electrons at nearest neighbor sites.
$\Delta (>0)$ denotes a kind of local on-site energy. 
Because of this term, energetically  electrons tend to
stay at A-sites. 
If all the electrons are on A-sites, then the system does not cost
energy due to $\Delta$. Therefore, we define this state as the perfect LT state.
Because of the difference of the degeneracy listed above, at high temperature, electrons tend to stay at B-sites.
We express the degree of high temperature (HT) state by the number of electrons at B-sites
\beq
\langle n_{\rm B}\rangle={\sum_{i \in {\rm B}}n_i\over N}.
\eeq
In the above choice of the degeneracy, $\langle n_{\rm B}\rangle$ is more than half 
at high temperatures.
In the present model, no spin-crossover transition occurs by the electron transfer. 
However, the degeneracy of spins still plays an important role, and this model
exhibits a first order phase transition.\cite{ct3}
In the mean field theory, the self-consistent equation for 
$q=\langle \sigma_{i\in {\rm B}}\rangle$
is given by
\beq
q=\tanh\left[\beta\left({z\varepsilon q\over 4}-{\Delta\over 4}+k_{\rm B}T\ln(5/3) \right)\right],
\eeq 
where $z$ is the number of the nearest neighbor sites and is 3 for the Honeycomb lattice.
We depict the typical temperature dependence of solution of self-consistent equation for 
$\langle n_{\rm B}\rangle=(q+1)/2$ in  Figs.~\ref{fig-CTnB} for  $\Delta=1$ and 10. 
There, we fix $\varepsilon$ to be 3.
In Fig.~\ref{fig-CTnB}(b), 
there are three solutions for $q$ at low temperatures. There, the largest and smallest ones give
stable solutions and the second one gives an unstable solution. By comparing the free energies
corresponding to them, we find the first order phase transition point, which is $T_0$.
In figures, the temperature where $\langle n_{\rm B}\rangle=0.5$ is shown by a dotted lines.
This temperature has the same physical meaning as to $T_0$ defined in (\ref{T0}) as will be
shown in the following subsections.  

Here we find again that there are cases where the change is smooth and discontinuous.
However, in this charge transfer transition, the metastable branch of HT state remains until $T=0$, 
i.e., HT state is metastable even at very low temperatures.

\subsection{Comparison}

In the two observations in the above SC and CT cases,
there are qualitatively different temperature dependences
of the free energy structure. 
Thus, it would be interesting to compare these two cases, and 
we will attempt to express this model (\ref{HamCT}) in the form of Wajnflasz model 
(\ref{W-model}).

First, we transform the variable $n_i$(= 0 or 1) to  $s_i$(= $-1$ or 1):
\beq
n_i={s_i+1\over 2}.
\eeq
Here the constraint ($\sum_in_i=N$) becomes
\beq
\sum_is_i=0.
\label{constraint-s}
\eeq
The Hamiltonian becomes
\beq
{\cal H}={\varepsilon\over 4}\sum_{<ij>}(s_i+1)(s_j+1)+{\Delta\over2}\sum_{i\in {\rm B}}(s_i+1).
\eeq
Making use of the constraint (\ref{constraint-s}) we can rewrite it as
\beq
{\cal H}={\varepsilon\over 4}\sum_{<ij>}(s_i+1)(s_j+1)
+{\Delta\over 4}\left(\sum_{i\in {\rm B}}(s_i+1)-\sum_{i\in {\rm A}}(s_j+1)\right).
\eeq 
In order to have a ferromagnetic model, we perform a local gauge transformation:
$\varepsilon \rightarrow -\varepsilon$, $s_i \rightarrow -S_i$ at A-site and 
$s_i \rightarrow S_i$ at B-site. Then we have
\beq
{\cal H}=-{\varepsilon\over 4}\sum_{<ij>}S_iS_j
+{\Delta\over 4}\left(\sum_{i\in{\rm A}}S_i+\sum_{i\in {\rm B}}S_i\right).
\eeq
Here we again use non-degenerate Ising variable $\sigma_i$.
It should be noted that the numbers of degeneracy at A- and B-sites are different.
However, 
using relations similar to (\ref{degeneracy}), 
the numbers of degeneracy 
($n_{\rm A1}$, $n_{\rm A0}$, $n_{\rm B1}$ and $n_{\rm B0}$) can be taken into account as
\beq
{\cal H}=-{\varepsilon\over 4}\sum_{<ij>}\sigma_i\sigma_j
+\left({\Delta\over 4}+{1\over2}k_{\rm B}T\ln{n_{\rm A1}\over n_{\rm A0}}\right)\sum_{i\in{\rm A}}\sigma_i
+\left({\Delta\over 4}-{1\over2}k_{\rm B}T\ln{n_{\rm B1}\over n_{\rm B0}}\right)\sum_{i\in {\rm B}}\sigma_i.
\eeq
Using the constraint 
\beq
\sum_{i\in{\rm A}}\sigma_i-\sum_{i\in {\rm B}}\sigma_i=0,
\label{constraint-S}
\eeq
we finally have
\beq
{\cal H}=-{\varepsilon\over 4}\sum_{<ij>}\sigma_i\sigma_j
+\left({\Delta\over 4}+{1\over 4}k_{\rm B}T\ln{g_{\rm A}\over g_{\rm B}}\right)\sum_{i}\sigma_i,
\eeq
where $g_{\rm A}=n_{\rm A1}/ n_{\rm A0}$ and $g_{\rm B}=n_{\rm B1}/ n_{\rm B0}$.
Now, we have the same form as that of Wajnflasz model (\ref{W-model}).
Therefore, it is proved that the model (\ref{HamCT}) is equivalent to the model (\ref{SCs}).

\subsection{$T_0$ for the CT transition }

In this formalism, we can make use of the relation (\ref{T0}) to distinguish the smooth and discontinuous changes.
Here, the transition temperature $T_0$ is given by 
\beq
k_{\rm B}T_0={\Delta\over \ln{{g_{\rm B}\over {g_{\rm A}}}}}.
\eeq
Substituting the numbers of degeneracy, we have 
\beq
k_{\rm B}T_0={\Delta\over \ln{5/6\over 1/2}}={\Delta\over\ln{5/3}}\simeq 1.958\Delta.
\eeq
In this subsection, we set $k_{B} =1$.

In the case  $\Delta=10$, the condition (\ref{smooth}) is satisfied. 
That is,
in  the honeycomb lattice, the critical temperature of the ferromagnetic
Ising model is given by $k_{B}T_{\rm CI}=3(\varepsilon /4)$ in the mean-field theory, and 
therefore, for $\varepsilon=3$,
\beq
T_{\rm CI}=0.75\times 3 < T_0=1.958\times 10,
\eeq
and thus $\langle n_{\rm B}\rangle$ changes smoothly, which is depicted in Fig.~\ref{fig-CTnB}(a).
On the other hand,
for $\Delta=1$,
\beq
T_{\rm CI}=0.75\times 3 > T_0=1.958\times 1,
\eeq
and the transition is of the first order, which is depicted in Fig.~\ref{fig-CTnB}(b).

\subsection{Monte Carlo study}

\begin{figure}
$$
\begin{array}{c}
\epsfxsize=7.0cm \epsfysize=6.5cm \epsfbox{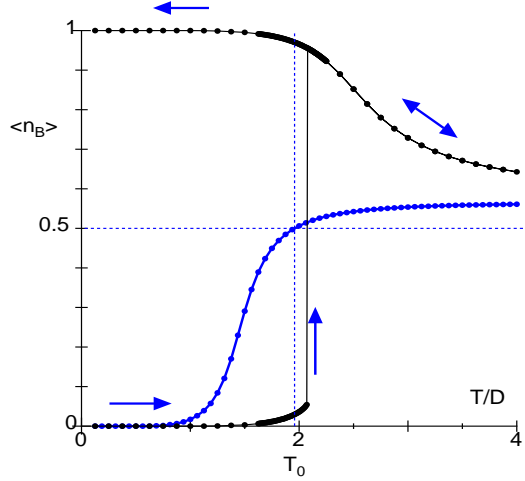}
\end{array}
$$
\caption{Temperature dependence of fraction of the HT state $\langle n_{\rm B}\rangle$ 
for $\varepsilon=3$ (blue dots) and 6 (black dots) studied by Monte Carlo
 simulations. Here, $\Delta =1$.}  
\label{fig-CTnBMC}
\end{figure}

It should be noted that, although in the above we study the model in the mean-field approximation,
the criterion 
whether the change of $\langle n_{\rm B}\rangle$ at $T_0$ is smooth or discontinuous,
i.e., the relation between $T_0$ and $T_{\rm CI}$, also holds in the exact treatment.
In the above treatment, the value of $T_0$ is given in the mean-field approximation
because we studied the model in the mean-field approximation. 
However, in the exact treatment, $T_0$ is given by the critical temperature of the ferromagnetic 
Ising model:
\beq
T_{\rm CI}={2\over\ln(2+\sqrt{3})}
{\varepsilon\over 4}\simeq1.52\times\left({\varepsilon\over 4}\right).
\eeq
In Fig.\ref{fig-CTnBMC}, we depict the temperature dependence of 
$\langle n_{\rm B}\rangle$ studied by
a Monte Carlo (MC) method for $\varepsilon=$ 3 and 6 with $\Delta = 1$. 
Because we study the exact short range model in MC, we find exact properties in MC
although it is numerically done.
For $\varepsilon =3$, 
\beq
T_{\rm CI}= 1.52\times(3/4) <T_0=1.958,
\eeq 
and then we indeed find that $\langle n_{\rm B}\rangle$
changes smoothly as plotted by the blue dots.
For discontinuous change of $\langle n_{\rm B}\rangle$, the relation 
$T_0 <T_{\rm CI}$, i.e., 
\beq
1.958 <  1.52\times(\varepsilon/4),
\eeq
must be held.
Therefore, in MC, $\varepsilon$ must be larger than $5.15 (=1.958\times 4/1.52)$ for the discontinuous change. 
The temperature dependence of  $\langle n_{\rm B}\rangle$ for $\varepsilon=6$ 
is also plotted in Fig.~\ref{fig-CTnBMC} with black dots,
where we find a discontinuous jump.
Here, we changed the value of $\varepsilon$ instead of $\Delta$ to have the same value of $T_0$
for the convenience to plot the data in a figure.

Monte Carlo simulations were performed on the honeycomb lattice of the size $32\times 32$. 
The simulation started at $T=0.125$ where $\langle n_{\rm B}\rangle\simeq 0$ in the
equilibrium state. The temperature was increased up to $T=4$. At each temperature, we
performed 20000 Monte Carlo steps (MCS) for a transient process and then
took data in the subsequence 100000MCS. 
Fluctuation of the data is smaller than the size of the dots.

We find that $\langle n_{\rm B}\rangle$ jumps at $T=2.12$ although the exact critical point is 
$T_0=1.958$ which is indicated by the dotted line where $\langle n_{\rm B}\rangle=0.5$.
This shift of the jump is due to the hysteresis phenomenon associated with 
the first order phase transition.

After $T=4$, the temperature was reduced. 
We find almost the same values of $\langle n_{\rm B}\rangle$ as those in the heating process
above $T=2.12$. However, it keeps a large value below this temperature, and stays 
in metastable HT state even below $T_0$. 
This metastability corresponds to the metastable solution found in the mean field theory.     
In the case $\varepsilon=3$, no hysteresis is found.

\begin{figure}
$$
\begin{array}{c}
\epsfxsize=7.0cm \epsfysize=6.5cm \epsfbox{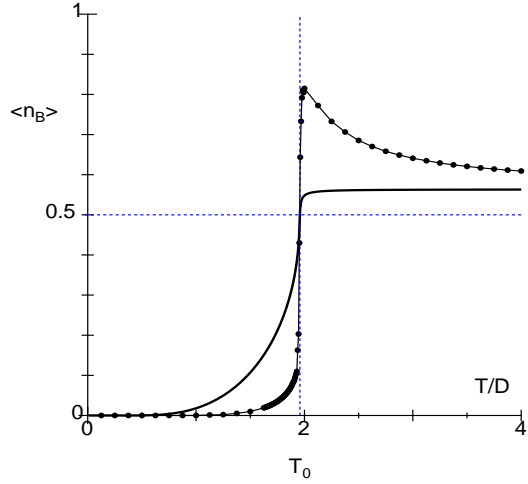}
\end{array}
$$
\caption{Temperature dependence of fraction of the HT state $\langle n_{\rm B}\rangle$ 
at the critical value of $\varepsilon$ studied by the mean-field theory 
$\varepsilon_{\rm C}=3/(4\ln{10/3})=2.61\cdots$ and by a Monte Carlo simulation
$\varepsilon_{\rm C}=1/(4\ln{10/3})=5.18\cdots $. Here, $\Delta =1$.}  
\label{fig-CTnBMCEC}
\end{figure}

In this way, we can know various properties of $\langle n_{\rm B}\rangle$
from the knowledge on the corresponding Ising model.
For example, at the critical value of $\varepsilon$, $\langle n_{\rm B}\rangle$
shows a second order phase transition where
\beq
\langle n_{\rm B}\rangle-{1\over 2}\propto |T-T_0|^{1\over \delta},\quad \delta=15
\eeq
and above the critical value of $\varepsilon$, 
$\langle n_{\rm B}\rangle$ changes discontinuously. 
The jump of $\langle n_{\rm B}\rangle$ is given by
\beq
\langle n_{\rm B}(T_0+0)- n_{\rm B}(T_0-0)\rangle\propto 
|\varepsilon-\varepsilon_c|^{\beta},\quad \beta={1\over 8},
\eeq
in the exact treatment and also in the Monte Carlo simulation. 
In the mean-field theory, on the other hand, the critical exponents are $\delta=3$ and $\beta=1/2$. 
In Fig.~\ref{fig-CTnBMCEC}, the temperature dependences of  $\langle n_{\rm B}\rangle$ 
at the critical value of $\varepsilon$ are plotted for both cases studied 
in the mean-field theory (by bold curve) and in MC (by circles). 
Here it should be noted that  $\langle n_{\rm B}\rangle$
shows a non-monotonic temperature dependence even in the case of the second order phase transition
in MC.

\section{Classification of temperature dependence of the high-spin fraction}

In the previous section, we have found the equivalence of the model of the SC
transition and the model of the CT transition. 
In the SC case, the HS state becomes unstable at a temperature $T'$ below the hysteresis loop. 
On the other hand, in the CT case the HT state remains metastable below $T_0$ until $T=0$.
These two observations suggest that there are qualitatively different temperature dependences
of the free energy structure. By studying these structures, we can reach a 
comprehensive understanding of phase transitions of systems of spin-crossover atoms.

Here we study the model of the type of (\ref{W-model}) with the variable $\sigma_i=\pm 1$
\beq
{\cal H}=-J\sum_{<ij>}\sigma_i\sigma_j+\left(D-{k_{\rm B}T\over 2}\ln{g}\right)\sum_i\sigma_i.
\eeq
As has been mentioned in the previous section,
the temperature $T_0$ where $\langle \sigma_i\rangle=0$ is given by (\ref{T0}).
Here, we define $D_{\rm C}$ as that the system exhibits a first-order phase transition
when $D< D_{\rm C}$:
\beq
D_{\rm C}={k_{\rm B}T_{\rm C}\over 2}\ln g.
\eeq  
Next, we consider the condition for the metastability of the HT state at the ground state.
By flipping one spin in the configuration where all the spins are $+1$, 
the system  gains the crystal field $-2D$, while it loses the exchange 
energy $2zJ$ where $z$ is the number of nearest neighbors.
Thus, if $D$ is smaller than the critical value:
\beq
D_{\rm CG}=zJ,
\eeq
the all up state is metastable at $T=0$.

\begin{figure}
$$
\begin{array}{c}
\epsfxsize=7.0cm \epsfysize=6.5cm \epsfbox{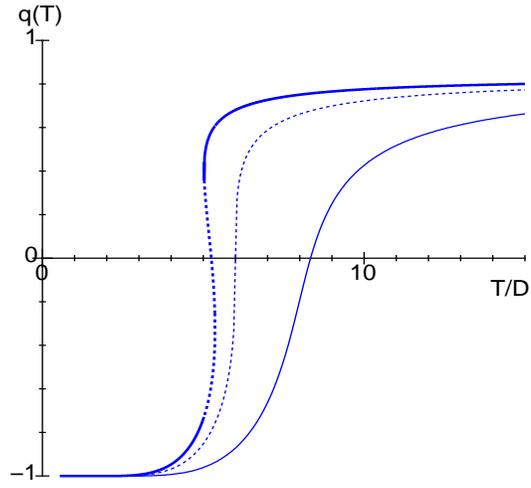}
\end{array}
$$
\caption{Temperature dependences of $q(T)$ are depicted for $D$ = 10.0 (thin solid curve),
$D=7.2$ (dashed curve), and $D=6.3$ (bold solid and dashed curve).
Here, $J=1$ and $\alpha=1.2$ $(g=e^{2.4})$.}
\label{fig-SCqHT}
\end{figure}

\begin{figure}
$$
\begin{array}{cc}
\epsfxsize=7.0cm \epsfysize=6.5cm \epsfbox{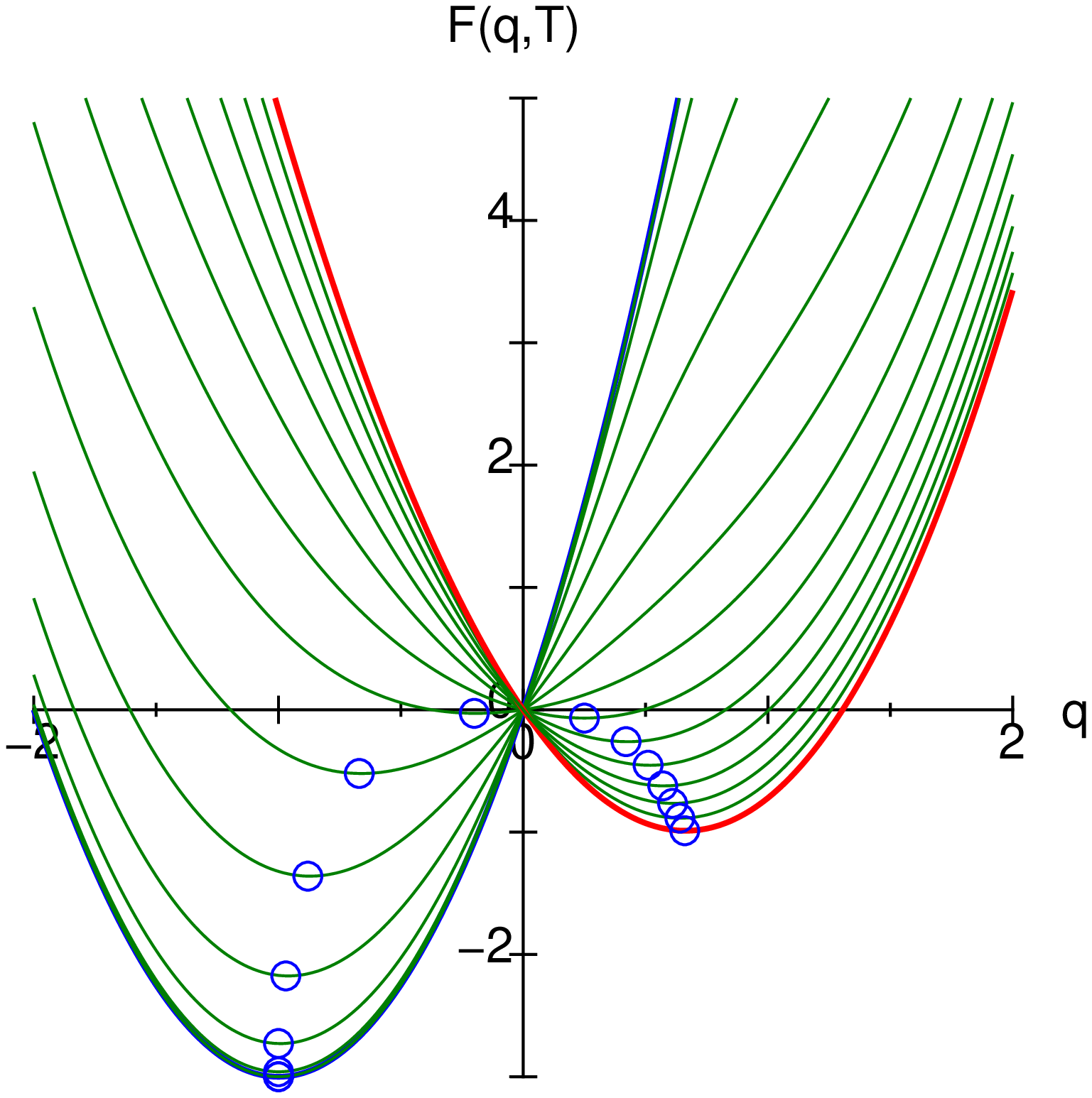}
&
\epsfxsize=7.0cm \epsfysize=6.5cm \epsfbox{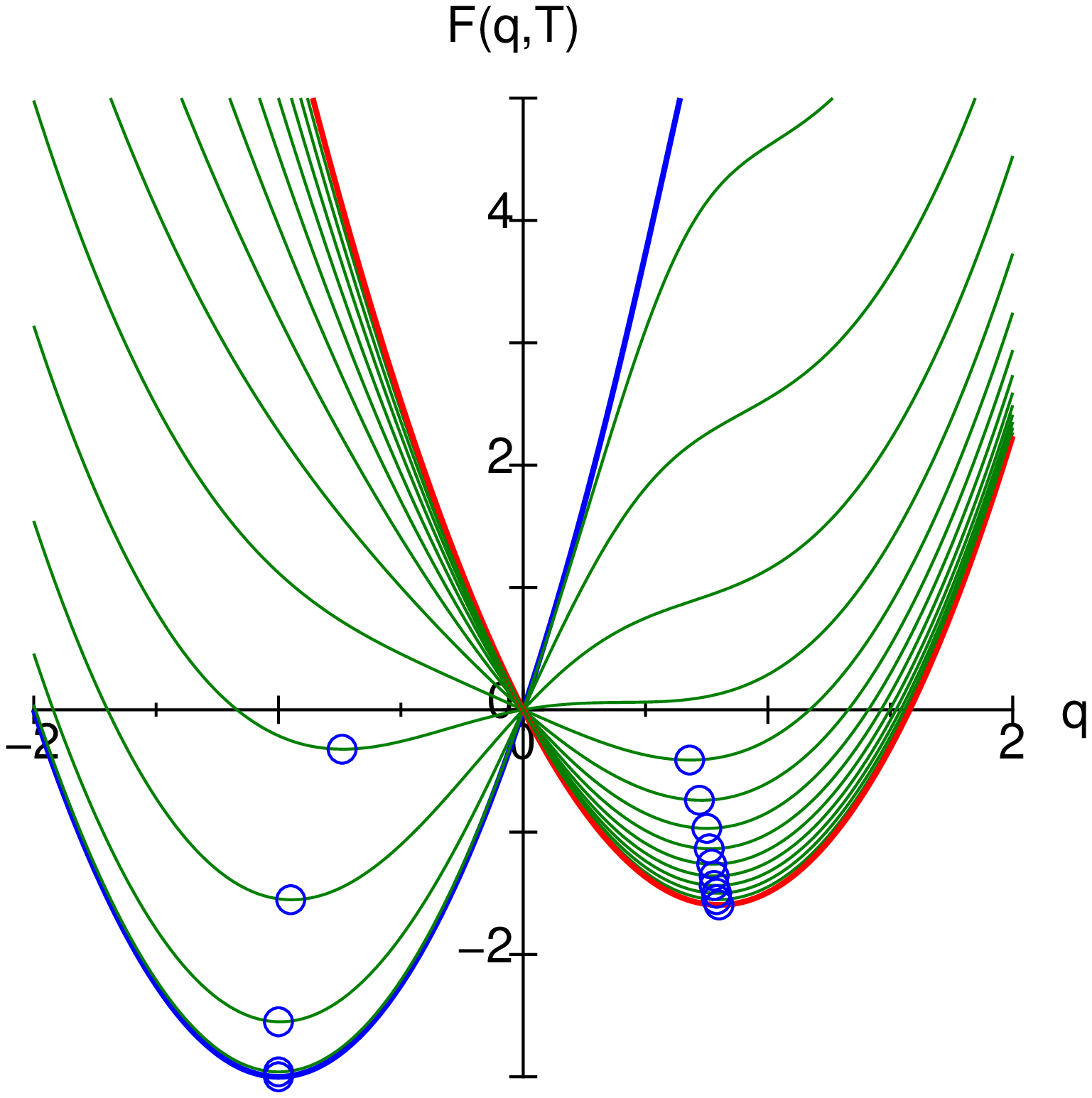}\\
{\rm (a)}&{\rm (b)}
\end{array}
$$
\caption{Temperature dependence of the free energy of functions of $q$ are depicted for
(a) $D$ = 10.0, and (b) $D=6.3$. The temperature changes by 1 from $T=$ 15 (red) to 1 (blue).
Here $J=1$ and $\alpha=1.2$ $(g=e^{2.4})$. The open circles indicate extreme points of the free energy.}
\label{FM-HT}
\end{figure}

Let us study temperature dependence of solution(s) 
\beq
q(T)=\langle \sigma_i\rangle
\eeq
of mean-field self-consistent equation:
\beq
q=\tanh\left[\beta\left(Jzq-D+{k_{\rm B}T\over 2}\ln g\right)\right].
\label{self-consistent}
\eeq
The free energy for the solution is given by 
\beq
F(q)={1\over 2}zJq^2-k_{\rm B}T\ln\left[2\cosh\left({Jzq-D\over k_{\rm B}T}+\alpha \right)\right].
\eeq 

Here, we adopt the following parameters 
\beq
J=1,\quad z=6,\quad  {\rm and }\quad \alpha\equiv{1\over2}\ln{g}=1.2,
\eeq
and here we take $J$ as a unit of the energy. In the followings,
we study the temperature dependence of $q(T)$ and its dependence on $D$. 

\subsection{Large $D$ behavior}

In Fig.~\ref{fig-SCqHT}, we show the temperature dependence of $q$ for $D=10$ by a thin solid curve
where we find a smooth change of $q$, and corresponding temperature dependence of
the free energy $F(q)-F(0)$ is depicted in Fig.~\ref{FM-HT}(a). 
In the present case $D_{\rm C}$ is 7.2, below which the system shows a first-order phase transition.
Temperature dependence of $q(T)$ for $D=7.2$ is depicted by a dotted curve.
The temperature dependence of $q$ for $D=6.3$ is depicted
by bold solid and dashed curve, which shows a first order phase transition.
The overhanging part is plotted by a bold dotted line.
This temperature dependence corresponds to the standard phenomena of HS-LS transition
shown in Fig.~\ref{fig-hysteresis}.
Corresponding temperature dependence of the free energy is depicted in Fig.~\ref{FM-HT}(b).
\begin{figure}
$$
\begin{array}{c}
\epsfxsize=7.0cm \epsfysize=6.5cm \epsfbox{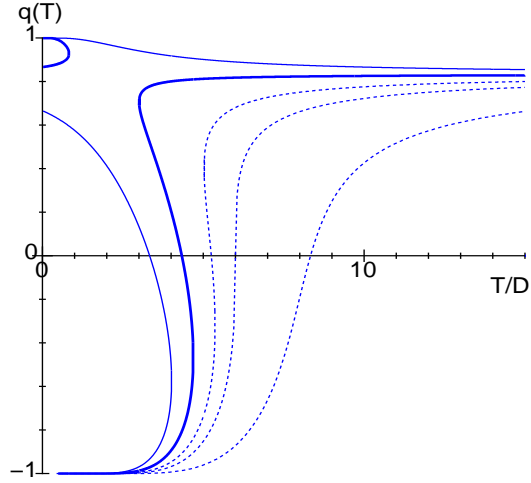}
\end{array}
$$
\caption{Temperature dependences of $q$ are depicted for $D$ = 5.2 (bold solid curve) 
and $D=4.0$ (thin solid curve). The dotted curves are $q(T)$ for $D=$ 10, 7.2, and 6.3 which are
depicted in Fig.~\ref{fig-SCqHT}.
Here, $J=1$ and $\alpha=1.2$ $(g=e^{2.4})$.}
\label{fig-SCqLT}
\end{figure}

\begin{figure}
$$
\begin{array}{cc}
\epsfxsize=7.0cm \epsfysize=6.5cm \epsfbox{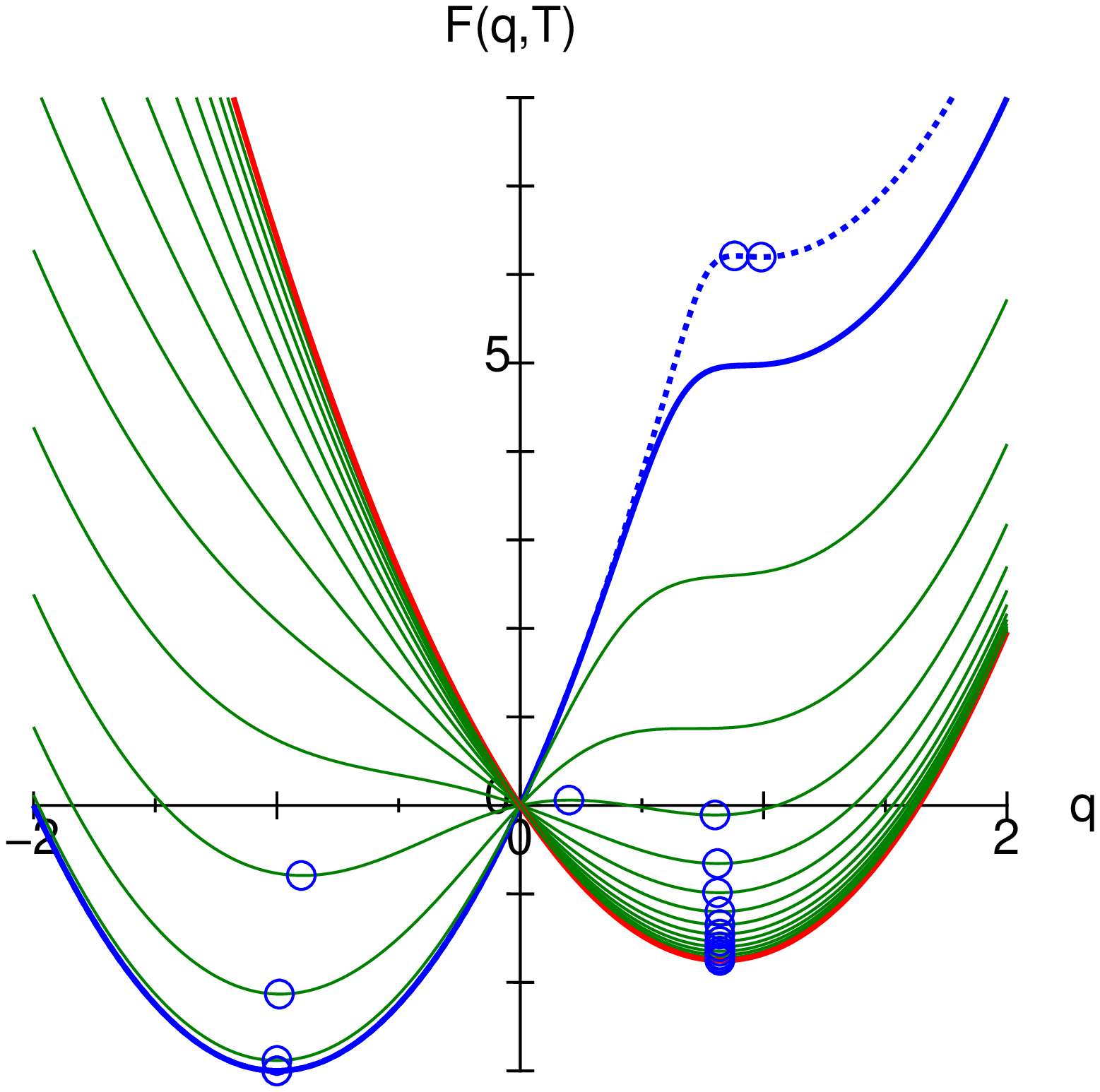}
&
\epsfxsize=7.0cm \epsfysize=6.5cm \epsfbox{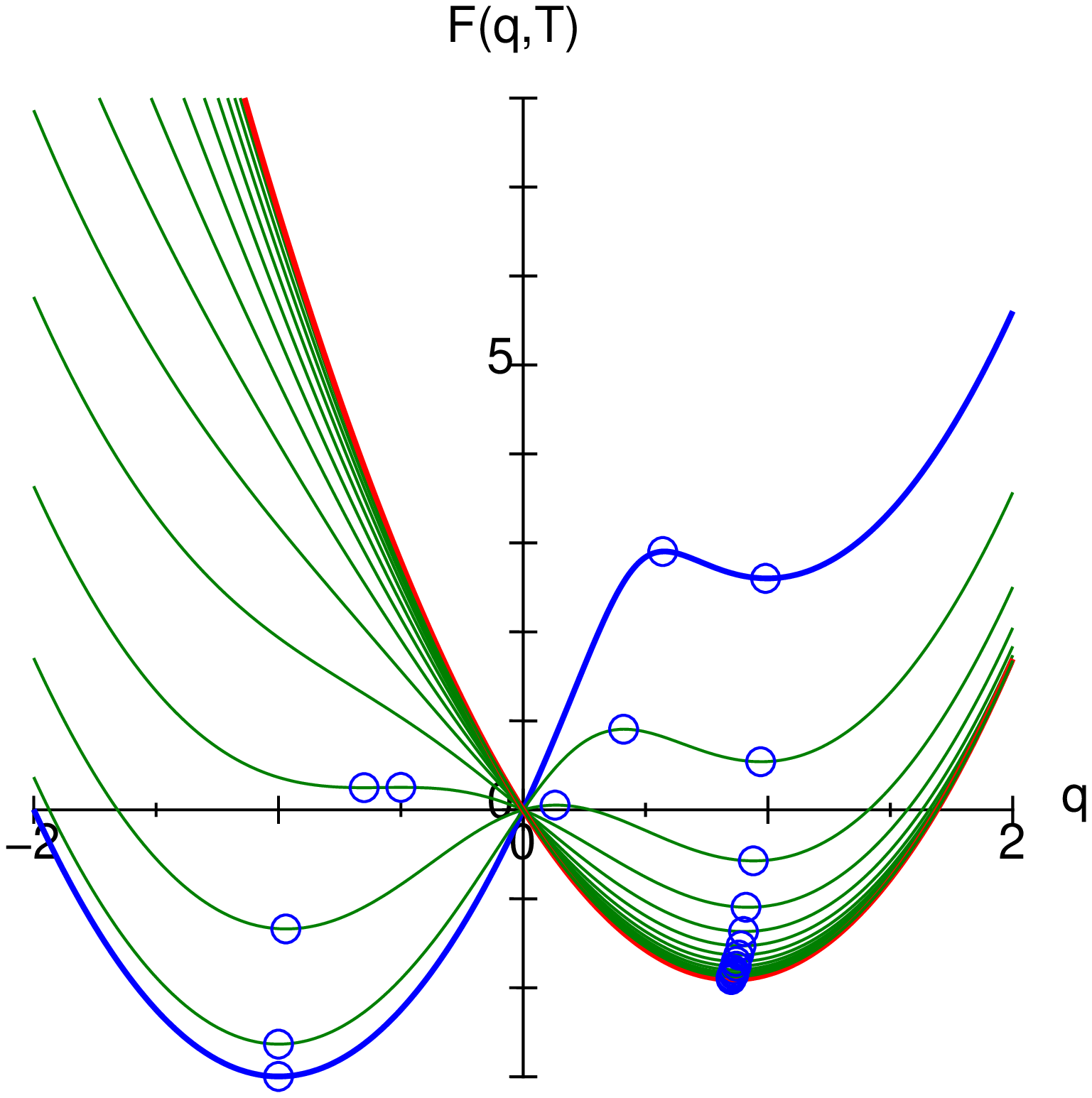}\\
{\rm (a)}&{\rm (b)}
\end{array}
$$
\caption{Temperature dependence of the free energy of functions of $q$ for
(a) $D$ = 5.2  and (b) $D=4.0$. The temperature changes by 1 from $T=$ 15 (red) to 1 (blue).
The data for $T=0.5$ is also plotted by a bold dotted blue curves.
Here, $J=1$ and $\alpha=1.2$ $(g=e^{2.4})$.}
\label{FM-LT}
\end{figure}

In typical spin-crossover transitions, 
the ratio of the degeneracy $g$ is considered to be rather large, and it causes
the value of $q$ in the high temperature limit 
\beq
q(\infty)=\tanh(\alpha)={g-1\over g+1}
\eeq
to be large. 

\subsection{Hysteresis branch in low temperature region for the intermediate $D$}

If we take the value of $D$ below $D_{\rm CG}$, we find a metastable branch
appears at low temperatures. 
In Fig.~\ref{fig-SCqLT}, we show the temperature dependence of $q$ for $D=5.2$ by
a bold curve,
and corresponding temperature dependence of the free energy in Fig.~\ref{FM-LT}(a).
Here we find the free energy has a local minimum of HS at low temperatures.
This low temperature metastable branch of HS may play an important role 
for the long-lived metastable state in Co-Fe PBA. 
In this case, the metastability of HS is intrinsic and 
if the initial state is set to be near the metastable point,
the value of $q$ is expected to move first to the metastable value and then 
relaxes to the LT state through a nucleation process. 
On the other hand, if there is no metastable branch 
at low temperatures as the case $D=6.3$,  
the long-lived HS state must be  purely due to the slow dynamics.
In this case,  
$q$ always moves to the 0. 
Checking this initial move of $q$, one may
find which case is realized. It would be an interesting problem to
determine that to which case individual materials belong.

\subsection{Small $D$ behavior}

\begin{figure}
$$
\begin{array}{c}
\epsfxsize=7.0cm \epsfysize=6.5cm \epsfbox{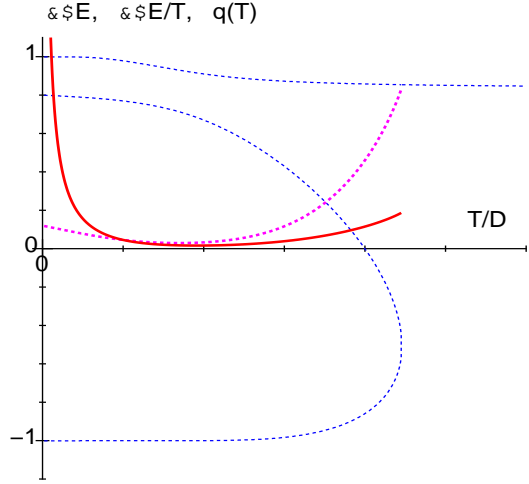}
\end{array}
$$
\caption{Temperature dependence of the barrier height between the metastable HT state and the stable LT state for
$D=4.8$. The temperature dependence of $q(T)$ is also plotted by thin dotted curves.
The energy barrier $\Delta E$ is depicted by the bold dotted curve, and 
$\Delta E/k_{\rm B}T$ is depicted by the bold  curve.
Here, $J=1$ and $\alpha=1.2$ $(g=e^{2.4})$.
}
\label{Barrier}
\end{figure}

If we reduce $D$ furthermore, e.g., $D=4$, HT state becomes metastable at all temperatures
as we depict in Fig.~\ref{fig-SCqLT} by a thin curve. The free energy change of this case is plotted in
Fig.~\ref{FM-LT}(b). 
This case corresponds to the dependence found in CT transition as shown in Fig.~\ref{fig-CTnB}.
For $D=4$, the high temperature branch connects to the metastable
HT branch at low temperatures, which is topologically different from those 
of large values of $D$. 
It should be noted that in this case the potential barrier between
the metastable HT state and the stable LT state shows a non-monotonic dependence,
and the metastable state easily relaxes to the stable LT state via a nucleation process
in the intermediate temperature region.
The dependence of the barrier is plotted in Fig.~\ref{Barrier}.

\subsection{Topological change of the structure of solutions}

\begin{figure}
$$
\begin{array}{c}
\epsfxsize=7.0cm \epsfysize=6.5cm \epsfbox{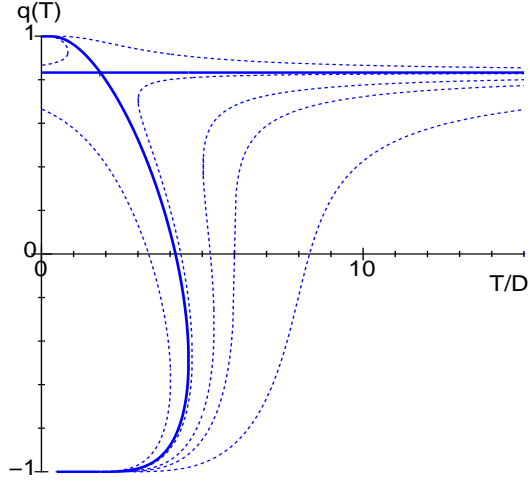}
\end{array}
$$
\caption{Temperature dependence of $q(T)$ for $D=D_{\rm CX}$ (solid curve).
The temperature dependence for the values studied previously are shown
 by dotted lines ($D = 10, 7.2, 6.3, 5.2$ and 4.0).
Here, $J=1$ and $\alpha=1.2$ $(g=e^{2.4})$.}
\label{fig-SCqCX}
\end{figure}

So far we found two regions of $D$.
In the large $D$ region the solution of HT state at high temperatures connects to
the LT state at low temperatures. On the other hand,
in the small $D$ region, the solution of HT state at high temperatures remains
until $T=0$. There, a LT solution exists separately.  
Thus, there is a critical value of $D$ between these two regions.
We depict $q(T)$ at the critical value of $D$ in Fig.~\ref{fig-SCqCX}. 
In the mean-field theory, $q(T)$ does not
depend on the temperature at this marginal value of $D$.
The values of $q(T)$ and the marginal $D$ are found in the following analysis.
If the following relations
\beq
q={D\over Jz}\quad {\rm  and} \quad q=\tanh\alpha, 
\eeq
hold, then from the self-consistent equation (\ref{self-consistent}), 
\beq
q=\tanh({Jzq-D\over k_{\rm B}T}+\alpha)
\eeq 
is always satisfied. From this relation, we find that the critical value
of $D$ is given by the equation: 
\beq
D_{\rm CX}=zJ\tanh\alpha
\eeq
and 
\beq
q=\tanh\alpha.
\eeq
Indeed the marginal value of $D$ is given by 
\beq
D_{\rm CX}=6\times\tanh(1.2)J=5.00193\cdots J.
\eeq

\subsection{Sequence of the types of structure}

Here we found three critical values of $D$.
Finally let us point out the fact that the isolated 
low temperature metastable branch of HS exists
for all the values of $\alpha$. That is, 
in order to have the metastable state at low temperature, 
$D$ must be smaller than $D_{\rm CG}$, i.e.,
\beq
D< zJ,
\eeq
and in order to avoid that the solution of HT exists at all the temperatures,
$D$ must satisfy the condition
\beq
D < D_{\rm CX}=zJ\tanh\alpha.
\eeq
Both conditions are compatible although the region of $D$ is rather narrow
for large $\alpha$.
In experiments, one may control the value of degeneracy by changing the pressure, etc.
It would be very interesting to find the qualitative change of temperature dependence 
of metastable states which is found here.

\section{Summary and Discussion}
 
In this paper, we studied models for phase transitions  
of systems consisting of spin-crossover atoms.
We pointed out that the apparently different models for the 
spin-crossover transition and for the charge-transfer transition are equivalent,
and we comprehensively studied 
structures of stable and metastable states of the unified model.
We found several qualitatively different 
ordering processes, and found critical values of $D$ between different
types of ordering processes. 
In particular, we found a metastable HS state at low temperatures
which exists separately from the high temperate HS state.
We find that a metastable HS state universally exists in the present type of models. 
This metastable HS state at low temperature would play an important in 
pumping process by photo-irradiation.
The present study will 
be useful to classify the metastability in various photomagnetic materials where
metastability will be investigated by photo-irradiation. 

In Co-Fe PBA and CT materials, the system shows a magnetic ordering in addition to the
spin structure change. There, we have to consider the magnetic interaction.
Effect of the magnetic interaction on the combined ordering process of
fraction ($q$) of HS and magnetic order ($m$) has become an interesting problem.
Here it should be noted that the magnetic state is always 
metastable but not equilibrium unless the magnitude of the 
interaction is of the order of $D$ in the models of SC.
If $D$ is so large that the magnetic state appears in equilibrium,
the change of HS-LS is smooth and the first order phase transition does not takes place.
The detailed properties of magnetic transition will be reported elsewhere.\cite{sct19,sct20,mag}
This feature of metastable magnetic state is compatible with the observation in Co-Fe PBA. 

In the CT system, however, magnetic ordered state appears as an equilibrium state
at a much lower temperature than the CT transition temperature. 
In order to explain this magnetic transition, we need to consider the special 
property of the CT system.
Indeed, in the CT system, magnetic moments exist on B-sites even at low temperatures.
Thus, if some mechanism exists to connect the magnetic moments, the magnetic order  
can be formed even in the low temperature configuration. One of the authors has proposed 
a mechanism that the thermal fluctuation of A-site could mediate the magnetic ordering 
between the B-sites.\cite{ct3} 
However, this mechanism has been found to be difficult at least
in two dimensional systems,\cite{ct4} and several alternate origins of magnetic ordering
are under investigation. In particular, 
it is possible that a quantum fluctuation helps the magnetic ordering 
which will be reported elsewhere.\cite{ct5}

\section*{Acknowledgements}
The present work is partially supported by
Grant-in-Aid from the Ministry of Education, Culture, Sports,
Science and Technology, and also by NAREGI Nanoscience Project, Ministry of
Education, Culture, Sports, Science and Technology, Japan.
The authors also thank the Supercomputer Center, Institute for
Solid State Physics, University of Tokyo for the facilities.

%

\end{document}